\documentclass{article}

\usepackage{arxiv}

\usepackage[utf8]{inputenc} 
\usepackage[T1]{fontenc}    
\usepackage{hyperref}       
\usepackage{url}            
\usepackage{booktabs}       
\usepackage{amsfonts}       
\usepackage{nicefrac}       
\usepackage{microtype}      
\usepackage{lipsum}
\usepackage{graphicx}
\graphicspath{ {./images/} }

\title{A benchmark analysis of saliency-based explainable deep learning methods for the morphological classification of radio galaxies}

\author{
 M.~T.~Atemkeng\\
  Department of Mathematics\\
  Rhodes University\\
  Grahamstown 6140 \\
  \texttt{m.atemkeng@ru.ac.za} \\
   \And
 C.~Chuma \\
  Department of Mathematics\\
  Rhodes University\\
  Grahamstown 6140 \\
  \texttt{caseymovex@gmail.com} \\
  \And
 S.~Zaza \\
  Department of Mathematics\\
  Rhodes University\\
  Grahamstown 6140 \\
  \texttt{s.zaza@ru.ac.za} \\
    \And
 C.~D.~Nunhokee \\
  International Centre for Radio Astronomy Research\\
  Curtin University\\
  Bentley WA 6102 \\
  \texttt{ridhima.nunhokee@curtin.edu.au} \\
    \And
 O.~M.~Smirnov \\
  Department of Physics and Electronics\\
  Rhodes University\\
  Grahamstown 6140 \\
  \texttt{o.smirnov@ru.ac.za} \\
}

\begin{document}
\maketitle
\begin{abstract}
This work proposes a saliency-based attribution framework to evaluate and compare 10 state-of-the-art explainability methods for deep learning models in astronomy, focusing on the classification of radio galaxy images. While previous work has primarily emphasized classification accuracy, we prioritize model interpretability. Qualitative assessments reveal that Score-CAM, Grad-CAM, and Grad-CAM++ consistently produce meaningful attribution maps, highlighting the brightest regions of FRI and FRII galaxies in alignment with known astrophysical features. In contrast, other methods often emphasize irrelevant or noisy areas, reducing their effectiveness. 
\end{abstract}


\section{Introduction}
Radio astronomy has entered an era of unprecedented data generation, driven by next-generation telescopes such as MeerKAT \cite{booth2012overview}
and the Square Kilometre Array (SKA) \cite{dewdney2009square}.
These instruments are designed to produce vast amounts of data, with the SKA expected to produce petabytes of data. While this data holds the potential for astronomical discoveries, it also presents significant challenges. Traditional methods of data analysis, which rely heavily on human expertise, are no longer scalable. The limited number of radio astronomy specialists and the time-consuming nature of manual analysis create bottlenecks that hinder the pace of discovery. Deep learning (DL) has emerged as transformative tools in radio astronomy, enabling the automation of complex tasks and addressing the challenges posed by the big data from modern telescopes. DL is applied to a growing array of tasks. For instance, DL algorithms have been successfully applied to tasks such as radio source classification, transient events detection, reconstruct high-resolution images from interferometric data, and identify faint signals in noisy datasets \cite{aniyan2017classifying, terris2023image}. 

Deep learning has offers transformative potential in radio astronomy due to its ability to uncover subtle patterns in complex datasets. However, the adoption of DL  is contingent on the robustness and interpretability of these models. Understanding how models arrive at their predictions is crucial for ensuring their reliability and for fostering trust among astronomers. Explainability techniques, such as saliency methods, have shown significant promise in addressing this challenge in other fields. These methods generate heatmaps that highlight the regions of an input image most influential in the model's feature extraction, providing insights into the classificatiion and/or detection mechanisms of DL models. 

In this work, we applied gradient-based saliency methods to infer where DL models extract features during the morphological classification of radio galaxies. This approach is particularly important in the classification of radio galaxies because the morphological features of radio sources, such as the presence of jets, lobes are key indicators of their physical properties and evolutionary stages \cite{fanaroff1974morphology}. By identifying which regions of an image the model prioritizes, we can validate whether the model is learning scientifically meaningful features, such as the extended emission from lobes or the compact core, rather than artifacts or noise. This not only enhances the interpretability of the model, but also ensures that its predictions align with established astrophysical knowledge. Furthermore, interpretable DL models can help astronomers uncover new insights into the diverse populations of radio galaxies, such as the transition between Fanaroff-Riley Type I (FRI) and Type II (FRII) sources or the identification of hybrid morphologies that challenge existing classification schemes.

\section{Image-based saliency}
We employ several well-known saliency methods; Vanilla Gradient calculates the output gradient in relation to the input image, highlighting the most influential pixels for predicting the target class \cite{mahankali2023beyond}. 
Integrated Gradients (IG) attributes the model's prediction to its input features by integrating the gradients along the path from a baseline to the input image \cite{lundstrom2022rigorous}. SmoothGrad IG enhances IG by averaging the gradients of multiple noisy versions of the input image, which helps reduce visual noise in the resulting saliency maps \cite{smilkov2017smoothgrad}. Guided Integrated Gradient (GIG) refines IG by guiding the gradients, resulting in less noisy and more interpretable saliency maps \cite{kapishnikov2021guided}. eXplanation with Ranked Area Integrals (XRAI) generates region-based attributions by ranking areas according to their contribution to the prediction, providing a more comprehensive view of the important regions. GradCAM uses the gradients of the target class flowing into the final convolutional layer to create a coarse localization map of significant areas in the image while GradCAM++ improves upon GradCAM by offering better localization by considering each neuron's importance in the last convolutional layer \cite{jamil2023advanced}. 
ScoreCAM relies on the confidence scores of the model to assess the importance of each activation map, which can result in more accurate and less noisy explanations \cite{wang2020score}.  
Vanilla Integrated Gradient uses the integrated gradient approach but applies it directly to the input in its unmodified form, providing attributions with minimal added complexity \cite{kapishnikov2021guided}. Blur Integrated Gradient applies a progressive blurring to the input image across integration steps, which helps emphasize key features while minimizing contributions from less relevant areas \cite{kapishnikov2021guided}.

These methods collectively offer a range of approaches to explain model decisions, from pixel-level gradients to region-based attributions, enhancing the interpretability and trustworthiness of DL predictions.

\section{Method and Framework}
The explainability process in a convolutional neural network (CNN) unfolds through a structured sequence of steps designed to identify the regions within an input image $X_i$ (from a dataset $D_m$ of $m$ samples) that most strongly influence the prediction of the network, ultimately visualized as a saliency map. This process begins by feeding an input image $X_i$ into the CNN, represented as $h(X_i, \phi)$, where $h(\cdot, \cdot)$ is the function that maps the input to the output through the layers of the network, and $\phi$ denotes the set of learnable parameters of the CNN. The model undergoes a series of transformations across convolutional, activation, pooling, and fully connected layers. These layers enable the network to progressively learn hierarchical features, from basic edges in the early layers to more complex, abstract representations in deeper layers. 
\begin{figure}
    \centering
     \includegraphics[width=0.5\columnwidth]{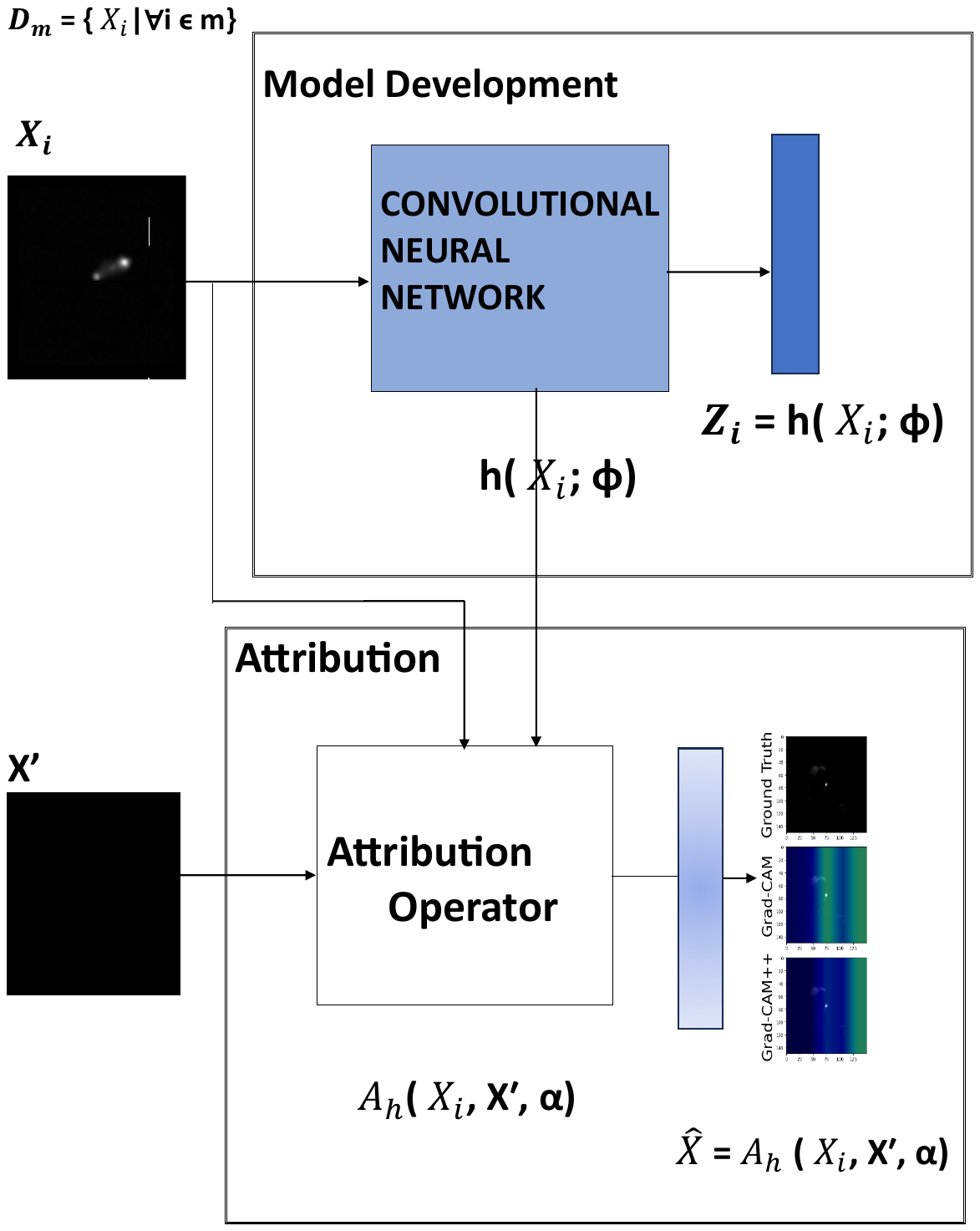}
    \caption{Overall pipeline to generate a saliency map.}
    \label{fig:enter-label}
\end{figure}

The forward pass through the CNN generates a prediction, and the attribution stage begins. In this stage, various explainability methods are employed to compute “attributions” that quantify the influence of each pixel or region of the image on the model's prediction for a specific class. This is done using an attribution operator $A_h$, which assigns importance scores to salient features $\hat{X}=A_h(X_i, X^\prime, \alpha)$, where $X^\prime$ is the output; a perturbed version of the input $X_i$ and $\alpha$  controls the weighting of features in the saliency map. Gradient-based methods compute attributions by calculating the gradient of the prediction score $z_i$ with respect to the input pixels $X_i$. This gradient indicates how small changes in the input pixels affect the prediction. SmoothGrad improves robustness by averaging gradients over multiple noisy versions of the input, while Grad-CAM computes gradients with respect to the activations of a specific convolutional layer, providing a coarse-grained saliency map. Perturbation-based methods, on the other hand, observe changes in the model’s output when specific regions of the input are masked or altered, offering an alternative perspective on feature importance. Figure \ref{fig:enter-label} illustrates the overall pipeline.
Once pixel-wise or region-based attributions are calculated, they are aggregated into a saliency map. This map visually emphasizes the areas that contributed the most significantly to the network’s decision, with intensity levels (often displayed as a heatmap) representing the importance of each region. Brighter colors typically indicate higher relevance, while other colors denote lesser importance. The final output overlays this saliency map on the original image, providing an intuitive visualization of the network's feature extraction. This enables one to assess whether the CNN focused on relevant features, offering valuable insight into the model’s interpretability and transparency. 

\begin{figure}
    \centering
     \includegraphics[width=1.0\columnwidth]{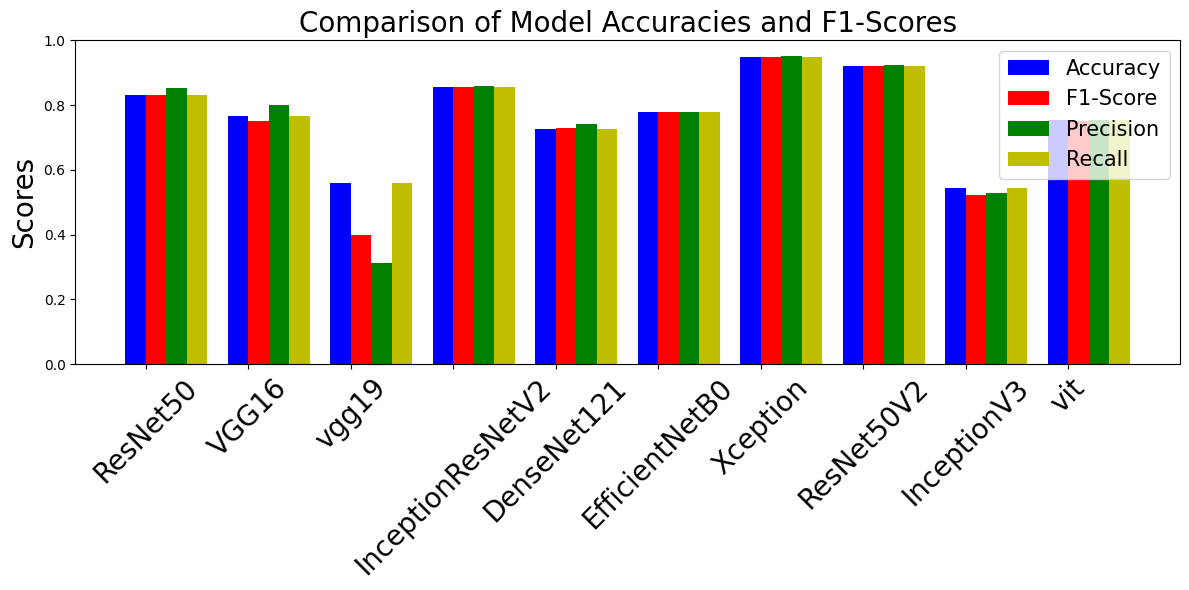}
    \caption{Performance classification comparison across several metrics.}
    \label{fig:enter-labelz}
 \end{figure}
\section{Experiment}
We use a subset of 770 radio images from the MiraBest dataset, a publicly available machine learning dataset containing radio-loud Active Galactic Nuclei (AGN) sourced from the NRAO VLA Sky Survey (NVSS) and the VLA FIRST Survey. This study focuses on classified sources with standard morphology, labeled as FRI and FRII  galaxies. These classifications are based on the well-established morphological distinction between FRI sources, which exhibit bright cores and fading jets, and FRII sources, which are characterized by edge-brightened lobes. To investigate the separability of these classes, we apply principal component analysis (PCA) to the dataset. The PCA reveals a clear separation between FRI and FRII radio sources in the reduced-dimensional space, demonstrating that the intrinsic morphological differences between these classes are well defined and can be captured by a DL model through its learned representations. 

The classification performance of our models is evaluated in Figure \ref{fig:enter-labelz}. The Xception model demonstrated superior performance across all evaluation metrics, achieving the highest F1 score, precision, and recall. 

\begin{figure}
    \centering
    \includegraphics[width=0.4\columnwidth]{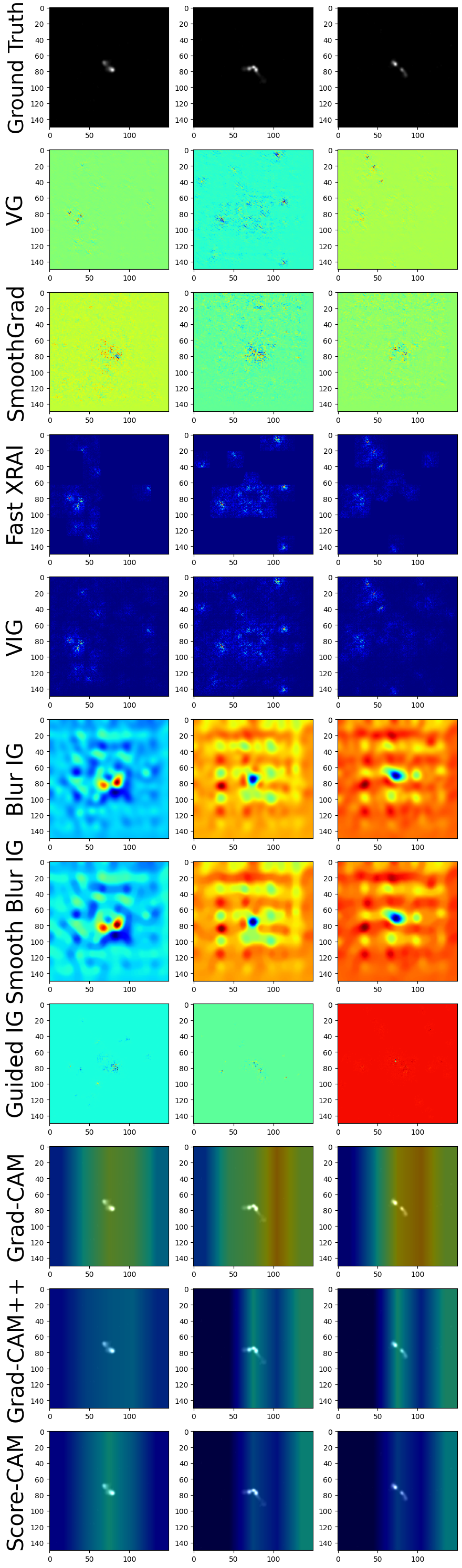}
    \caption{Benchmark of explainability methods applied to three FRI images classified by Xception. The first row shows the original FRI images, while the subsequent rows visualize where Xception extracts features for classification.}
    \label{fig:enter-label22}
\end{figure}

\begin{figure}
    \centering
\includegraphics[width=0.4\columnwidth]{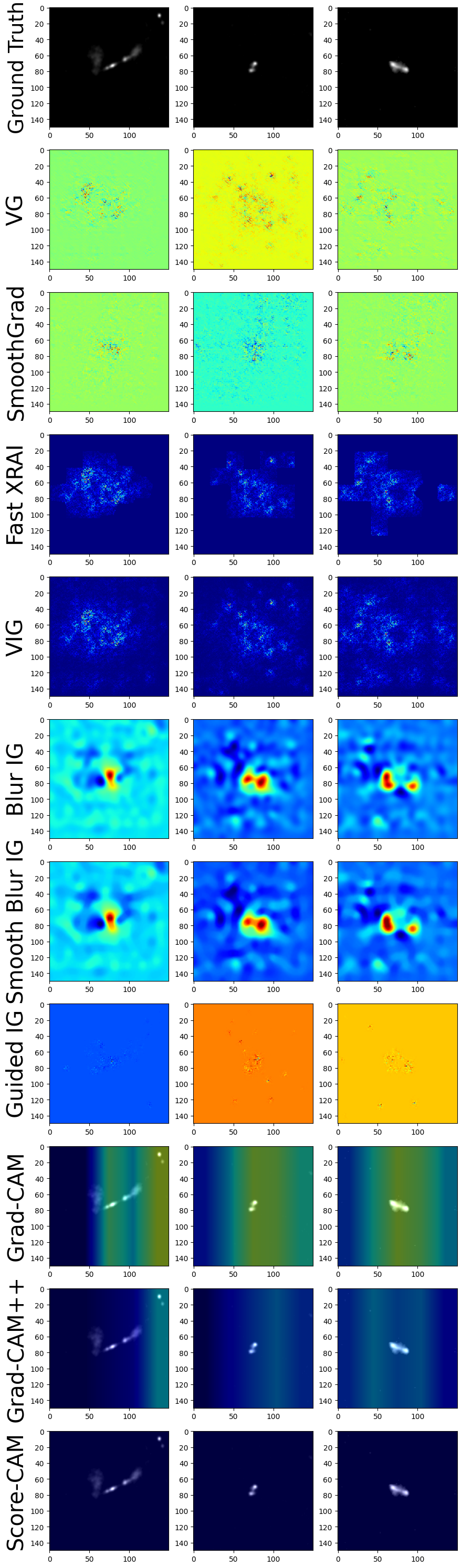}
    \caption{Benchmark of explainability methods applied to three FRII images classified by Xception. The first row shows the original FRI images, while the subsequent rows visualize where Xception extracts features for classification.}
    \label{fig:enter-label11}   
\end{figure}

Figures \ref{fig:enter-label22} and \ref{fig:enter-label11} visualize the feature attributions for three radio galaxy images from the test set, using a benchmark of explainability methods applied to the Xception model. The attribution maps highlight the regions of the input images that significantly influence the model predictions for FRI and FRII galaxies. The upper row displays the input images without attribution maps, while the subsequent rows show the attribution maps generated by each method. From visual inspection, Score-CAM, Grad-CAM, and Grad-CAM++ outperform the other methods, with Score-CAM producing the most precise and interpretable attribution maps, followed by Grad-CAM++. For FRI galaxies, these methods focus on the central region, which aligns with the astrophysical characteristic that FRI sources exhibit their brightest emission within the inner half of the source radius. For FRII galaxies, the methods emphasize the outer regions, consistent with the edge-brightened lobes that define this class.

Other methods, such as Vanilla Gradient and Smooth Gradient, produce attribution maps that are heavily influenced by noise around the image, reducing their interpretability. Similarly, XRAI and Fast XRAI 30\% generate comparable maps, although XRAI performs slightly better by being less affected by noise compared to Fast XRAI 30\%. Blur IG and Smooth Blur IG also exhibit similar attribution maps, with both methods showing high sensitivity to noise, which negatively impacts their ability to pinpoint meaningful regions. Most methods, apart from Score-CAM, Grad-CAM, and Grad-CAM++, tend to focus broadly on entire columns of pixels rather than specific regions, making it difficult to discern which features the model prioritized during the classification. This lack of specificity underscores the importance of selecting explainability methods that align with the scientific context, as methods such as Score-CAM and Grad-CAM++ provide clearer insight into the focus of the model, allowing better validation of its predictions against known astrophysical properties.

\section{Conclusions}
This work introduces a saliency-based attribution framework designed to evaluate and compare 10 state-of-the-art saliency methods, with the goal of improving the interpretability of DL models for classifying radio galaxy images. Unlike previous studies that predominantly prioritized classification accuracy, our work shifts the focus to the interpretability of the models. Through qualitative analysis, we find that Score-CAM, Grad-CAM, and Grad-CAM++ consistently generate interpretable and meaningful attribution maps for both FRI and FRII galaxy images. These methods effectively highlight the brightest regions of the galaxies, which correspond to known astrophysical features, despite some background noise. In comparison, other methods tend to focus on irrelevant or noisy areas, diminishing their utility in explaining radio galaxy classifications. 

Future work should incorporate quantitative metrics \cite{brima2024saliency} alongside qualitative assessments to further refine and validate these methods for radio galaxies classification.

\bibliographystyle{unsrt}

\bibliography{RSL-template}
\end{document}